\documentclass{iau}
\usepackage{graphicx}

\title[Polarization observations of large supernova remnants] 
{Radio polarization observations of large supernova remnants at $\lambda$6~cm
} 

\author[J. L. Han et al.]  
{J.~L. Han$^1$, X.~Y. Gao$^1$, X.~H. Sun$^{1,2}$, W. Reich$^2$, L. Xiao$^1$, P. Reich$^2$, J.~W. Xu$^1$, W.~B. Shi$^{1}$, E. F\"urst$^2$, \and R. Wielebinski$^2$
}

\affiliation{1. National Astronomical Observatories, Chinese Academy
  of Sciences, Jia-20 Datun Road, Chaoyang District, Beijing 100012,
  China.  hjl@nao.cas.cn \\ 
2. Max-Planck-Institut f\"{u}r  Radioastronomie, Auf dem H\"{u}gel 69, 53121 Bonn, Germany}

\pubyear{2014}
\volume{296}  
\pagerange{202--209}
\jname{\mbox{Supernova environmental impact}}
\editors{Dick McCray \& Alak Ray, ed.}
\begin{document}

\maketitle

\begin{abstract}
We have observed 79 supernova remnants (SNRs) with the Urumqi 25~m
telescope at $\lambda$6~cm during the Sino-German $\lambda$6~cm
polarization survey of the Galactic plane. We measured flux densities
of SNRs at $\lambda$6~cm, some of which are the first ever measured or
the measurements at the highest frequency, so that we can determine or
improve spectra of SNRs. Our observations have ruled out spectral
breaks or spectral flattening that were suggested for a few SNRs, and
confirmed the spectral break of S147. By combining our $\lambda$6~cm
maps with $\lambda$11~cm and $\lambda$21~cm maps from the Effelsberg
100~m telescope, we calculated the spectral index maps of several
large SNRs. For many remnants we obtained for the first time
polarization images, which show the intrinsic magnetic field
structures at $\lambda$6~cm. We disapproved three objects as being
SNRs, OA184, G192.8$-$1.1 and G16.8$-$1.1, which show a thermal
spectrum and no polarization. We have discovered two large supernova
remnants, G178.2$-$4.2 and G25.1$-$2.3., in the survey maps.
\end{abstract}

\firstsection 

\section{Introduction}

Radio observations of supernova remnants (SNRs) probe two aspects of
their physics. One aspect is particle acceleration and synchrotron
radiation. Radio images show the surface brightness distribution of
remnants, which in general is composed of diffuse emission and of
filamentary or shell-like emission. The particles were accelerated in
the shock-front of a SNR and radiate in the filamentary area. After
some time, the aged particles diffuse away from the shock area and
radiate over a much wider area, producing diffuse emission. The
filamentary emission produced by shock-accelerated particles shows a
power-law spectrum with an index of typically $\alpha=-0.4$ to $-0.5$
($S\sim \nu^{\alpha}$). This is for most of observed SNRs in adiabatic
expansion phase. Particles with high energies lose their energy faster
than those with low energies, so that the aged particles produce
extended emission with a steeper spectrum. Observations with adequate
resolution should be able to distinguish filamentary emission from
diffuse emission. To reveal the process of particle acceleration and
radiation, multiband observations are needed to make images of the
spectral index distribution.
The second aspect of SNR physics that can be probed with radio
observations is magnetic fields. Supernova explosions not only
accelerate particles, but also compress the surrounding medium by
their shocks. The magnetic fields penetrating the interstellar medium
are therefore compressed too, and then act as the agent for
accelerated particles to produce synchrotron radiation. Radio
polarization observations can probe the magnetic field structure of
SNRs. It has been found that young remnants have a radial field
structure, while old remnants have a tangential field structure
(e.g. \cite{fr04}). Note, however, that the foreground Faraday
rotation must be discounted to get the intrinsic polarization angles
of radio emission, so that the intrinsic magnetic field structure of
SNRs can be revealed. 

To address these two aspects of physics radio observations of SNRs
need 1) adequate angular resolution so that remnants can be resolved;
2) multiband intensity measurements so that an image of the spectral
index distribution can be calculated; 3) multiband polarization
measurements, so that Faraday rotation in the foreground can be
corrected, and the intrinsic magnetic field orientation can be figured
out. In addition, high enough observing frequencies, e.g. up to
$\lambda$6~cm, should be selected so that radio images of a SNR do not
get confused by fluctuating Galactic radio diffuse
emission. Observations at shorter wavelengths suffer from less
depolarization, and Faraday rotation changes position angles by only a
small amount.

There are many apparently large SNRs within 2-3~kpc distance in the
Galaxy, which have been objects for X-ray, Gamma-ray and optical
observations. However, their radio images at high radio frequencies
are extremely difficult to obtain. The huge apparent size of these SNRs,
a few degrees in general, is too large for synthesis telescopes due to
their limited field of view and their insensitivity to extended
emission, for a single large dish telescope, such as the Effelsberg
telescope, because of its small beam at high frequencies. We scanned
many large SNRs at $\lambda$6~cm to obtain polarization maps with the
single band polarization system of the Urumqi 25~m radio telescope of
Xinjiang Observatory, which we used for the Sino-German $\lambda$6~cm 
polarization survey of the Galactic plane. An excellent receiver was
constructed at the MPIfR and installed at the telescope in August 2004,
which has a very good stability for long scan-observations.

In this invited talk, we introduce the main results of radio
polarization observations of SNRs at $\lambda$6~cm by using the small
Urumqi 25~m telescope. Combining our data with $\lambda$21~cm
and $\lambda$11~cm observations made with the Effelsberg 100~m
telescope, we obtained many new results on SNR spectra and
polarization.

\section{Observational results}

Since August 2004, we have used the Urumqi 25~m radio telescope for
the Sino-German 6cm polarization survey of the Galactic plane in the
region of $10^{\circ}\leq l \leq 230^{\circ}$ and $|b| \leq 5^{\circ}$
(\cite{shr+07}, \cite{grh+10}, \cite{xhr+11}, \cite{srh+11a}). We
divided the survey region into many patches, and scanned them in both
Galactic longitude $l$ and latitude $b$. Regular observations of 3C286
or 3C295 are made for calibration purposes.  During the data
processing, we have to carefully fit the baselines, remove obvious
interference, and suppress scanning effects. The final radio maps at
$\lambda$6~cm were made in total power $I$ and linear polarization,
Stokes parameters $Q$ and $U$. The latter two maps were then combined
to calculate polarized intensity $PI$ and polarization angle $PA$
maps. The survey maps show more frequent fluctuations towards smaller
Galactic longitudes in the inner region (\cite{srh+11a},
\cite{xhr+11}) and the Cygnus region, which leads to high confusion
for SNRs there. Data for SNRs located in the survey region are
extracted from the survey, for some of them additional observations
were added for higher sensitivity. SNRs outside the survey region were
observed separately.

\begin{figure}[bt]
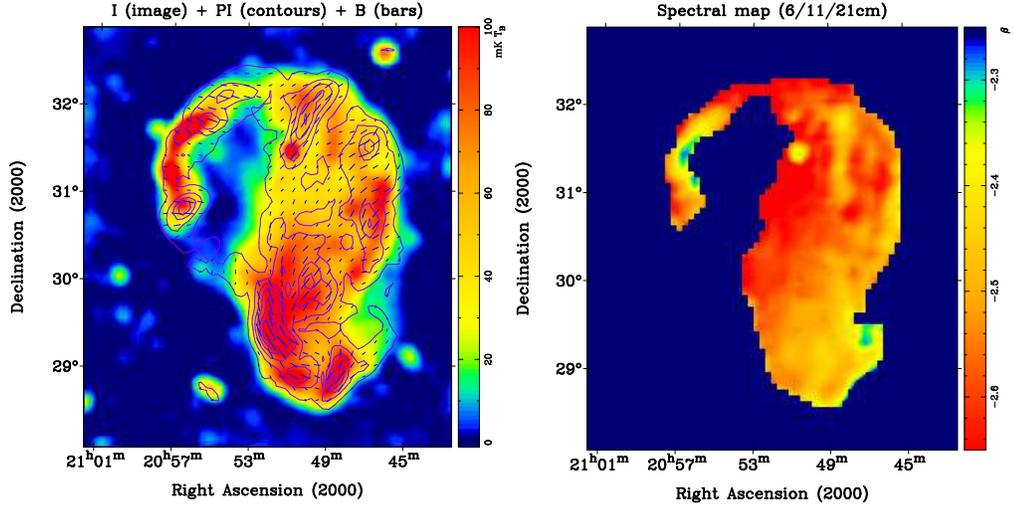

\centering
\includegraphics[angle=270,width=0.49\textwidth]{han_fig1a.ps}
\includegraphics[angle=270,width=0.49\textwidth]{han_fig1b.ps}
\caption{The first polarization map {\it (left)} was made for the
  Cygnus loop using the $\lambda$6~cm system. Combining our map
  with $\lambda$11~cm and $\lambda$21~cm maps from Effelsberg
  observations, we calculated a spectral index map {\it (right)}.
\label{cygnus}
}
\end{figure}
\subsection{The first polarization image of the $\lambda$6~cm system: the Cygnus loop}

The first polarization observations of the 6~cm system were made
towards the Cygnus loop (Fig.~\ref{cygnus}). This large SNR has a size
of $4^{\circ}\times3^{\circ}$. The polarization vector distribution
shows two polarization shells with different properties
(\cite{srh+06}), which supports the idea that the Cygnus Loop consists
of two SNRs, as suggested by \cite{ury+02}. Analysing the
$\lambda$6~cm map together with $\lambda$11~cm and $\lambda$21~cm data
from observations with the Effelsberg 100~m telescope, we got a
spectral index map (Fig.~\ref{cygnus}). Steep spectra are seen in the
central part of the SNR.

\begin{figure}[bt]
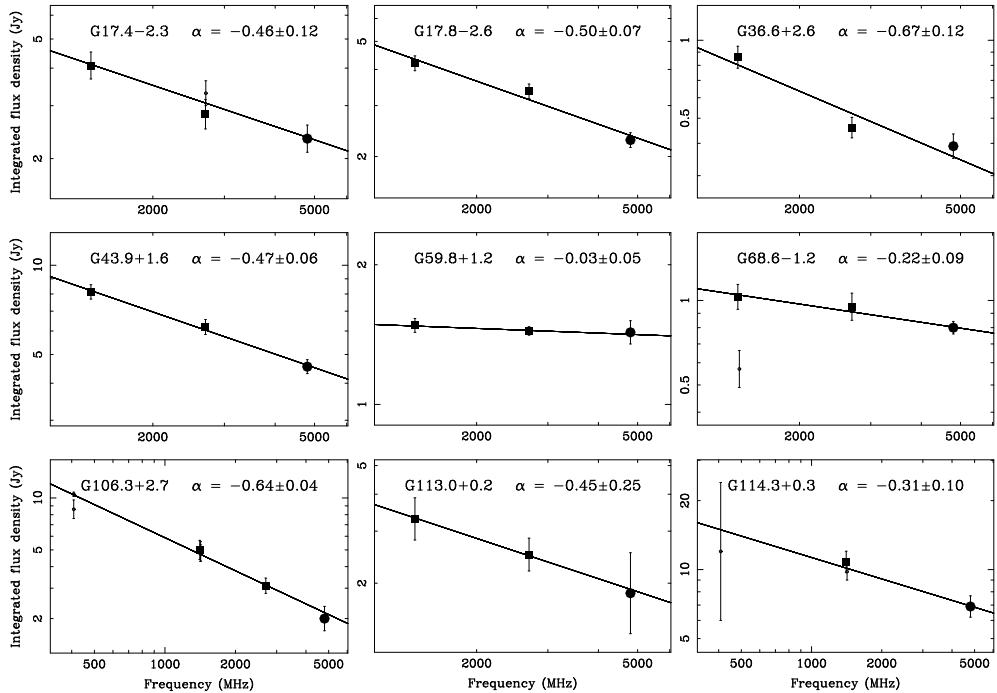

\centering
\resizebox{0.335\textwidth}{!}{\includegraphics[angle=-90]{han_fig2a.ps}}
\resizebox{0.31\textwidth}{!}{\includegraphics[angle=-90]{han_fig2b.ps}}
\resizebox{0.31\textwidth}{!}{\includegraphics[angle=-90]{han_fig2c.ps}} \\[2mm]
\resizebox{0.335\textwidth}{!}{\includegraphics[angle=-90]{han_fig2d.ps}}
\resizebox{0.31\textwidth}{!}{\includegraphics[angle=-90]{han_fig2e.ps}}
\resizebox{0.31\textwidth}{!}{\includegraphics[angle=-90]{han_fig2f.ps}}\\[2mm]
\resizebox{0.335\textwidth}{!}{\includegraphics[angle=-90]{han_fig2g.ps}}
\resizebox{0.31\textwidth}{!}{\includegraphics[angle=-90]{han_fig2h.ps}}
\resizebox{0.31\textwidth}{!}{\includegraphics[angle=-90]{han_fig2i.ps}}
\caption{Examples of newly determined flux densities (black symbols) at
  $\lambda$6~cm, $\lambda$11~cm, and $\lambda$21~cm to obtain spectra
  of SNRs (\cite{srr+11b}, \cite{ghr+11a}).
\label{newspec}
}
\end{figure}
\subsection{New flux density measurements for integrated spectra}

From our survey maps or separate observations at
$\lambda$6~cm we have measured the integrated flux densities of SNRs
(\cite{srr+11b}, \cite{ghr+11a}). For some SNRs, we also get new
measurements of integrated flux densities from the Effelsberg
$\lambda$11~cm and $\lambda$21~cm maps. Using these measurements,
together with measurements at other wavelengths from the literature,
we determined or improved integrated spectra of many SNRs
(Fig.~\ref{newspec}). For the SNRs G15.1$-$1.6, G16.2$-$2.7,
G16.4$-$0.5, G17.4$-$2.3, G17.8$-$2.6, G20.4 +0.1, G36.6+2.6, G43.9
+1.6, G53.6$-$2.2, G55.7 +3.4, G59.8+1.2, G65.1+0.6, G68.6$-$1.2,
G69.0 +2.7 (CTB 80), G93.7$-$0.2, G113.0+0.2, and G114.3+0.3, the
spectra have been significantly improved (\cite{srr+11b},
\cite{ghr+11a}).

\begin{figure}[b]
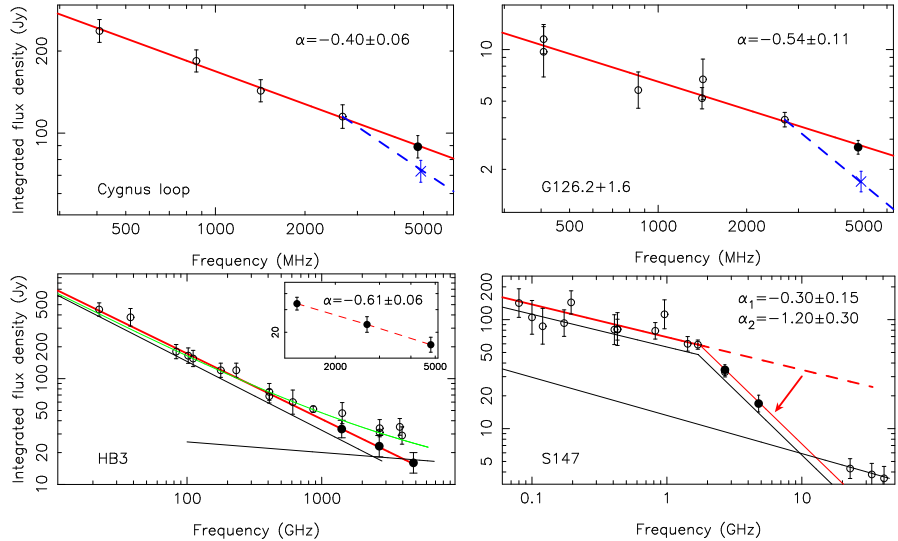

\centering
\includegraphics[angle=-90,width=0.44\textwidth]{han_fig3a.ps} \hspace{1mm}
\includegraphics[angle=-90,width=0.41\textwidth]{han_fig3b.ps} \\
\includegraphics[angle=-90,width=0.44\textwidth]{han_fig3c.ps} \hspace{1mm}
\includegraphics[angle=-90,width=0.41\textwidth]{han_fig3d.ps} 
\caption{New integrated flux densities at $\lambda$6~cm and
  $\lambda$11~cm and $\lambda$21~cm (black dots) rule out a spectral
  break previously proposed for the Cygnus loop and G126.2+1.6 and
  spectral flattening for HB3. We confirm the spectral break for S147.
\label{spec}
}
\end{figure}
\subsection{Spectral break or flattening}

A spectral break was suggested for several SNRs based on available
integrated flux densities in literature (see plots in
\cite{srr+11b}). SNR G74.9+1.2 seems to be a solid case which has a
break at about 10~GHz. SNR G31.9+0.0 has a flat spectral below a few
hundred MHz, probably due to the low frequency absorption. Flux
density data of SNR G21.5$-$0.9 and G69.0+2.7 (\cite{ghr+11a}) have
very large uncertainties for claiming a spectral break. SNR G27.8+0.6
may have a break, but to verify this requires more and better high
frequency data.

We disprove three claims of a possible spectral break or flattening by
using our new measurements of integrated flux densities at
$\lambda$6~cm.
The first one is the Cygnus loop (\cite{srh+06}). The small flux
density at 5~GHz from previous observations (\cite{kb72}) suggests
a spectral break above 2.7~GHz (\cite{uryf04}). Our new 
integrated flux density at $\lambda$6~cm rules out such a 
spectral steepening (see Fig.~\ref{spec}).
The second case was G126.2+1.6. \cite{tl06} suggested a spectral
break at about 1.5~GHz based on flux densities from the
literature with low accuracy. Our new measurement at $\lambda$6~cm
is consistent with a single power-law for the radio spectrum
(\cite{shr+07}).
The third case is HB3. \cite{upl07} claimed a spectral flattening
above 2~GHz as an indication for radio thermal bremsstrahlung emission
from a thin shell enclosing HB3. We obtained flux densities of HB3 at
$\lambda$6~cm, $\lambda$11~cm and $\lambda$21~cm for the region that
is not confused by the nearby strong HII region W3, and found that
these measurements are consistent with a single power law
(\cite{shg+08}).

We confirmed the case for the spectral break of S147, which is a
large, faint, shell-type SNR. Previous observations of S147 at 1648
MHz and 2700 MHz (\cite{kafh80}) and the southern part of the SNR at
4995 MHz (\cite{sfh80}) suggested a spectral break near 1.5 GHz, with
a flat spectrum at lower frequencies and a steep spectrum at higher
frequencies. The break needs to be confirmed because the previous
observations at 4995 MHz covered only the southern part of the
remnant.  Our new measurements cover the entire SNR with the Urumqi
25~m radio telescope at $\lambda$6~cm and new more sensitive
Effelsberg 100~m radio telescope at $\lambda$11~cm (\cite{xfrh08}),
and confirm a spectral break at $\sim$1.5 GHz. The spectral break is
caused by the combination of a diffuse emission component with a steep
spectrum and filamentary emission with a flat spectrum, which could be
traced up to 40~GHz including WMAP data (see Fig.~\ref{spec}).

\begin{figure}[tb]
\includegraphics[width=0.32\textwidth]{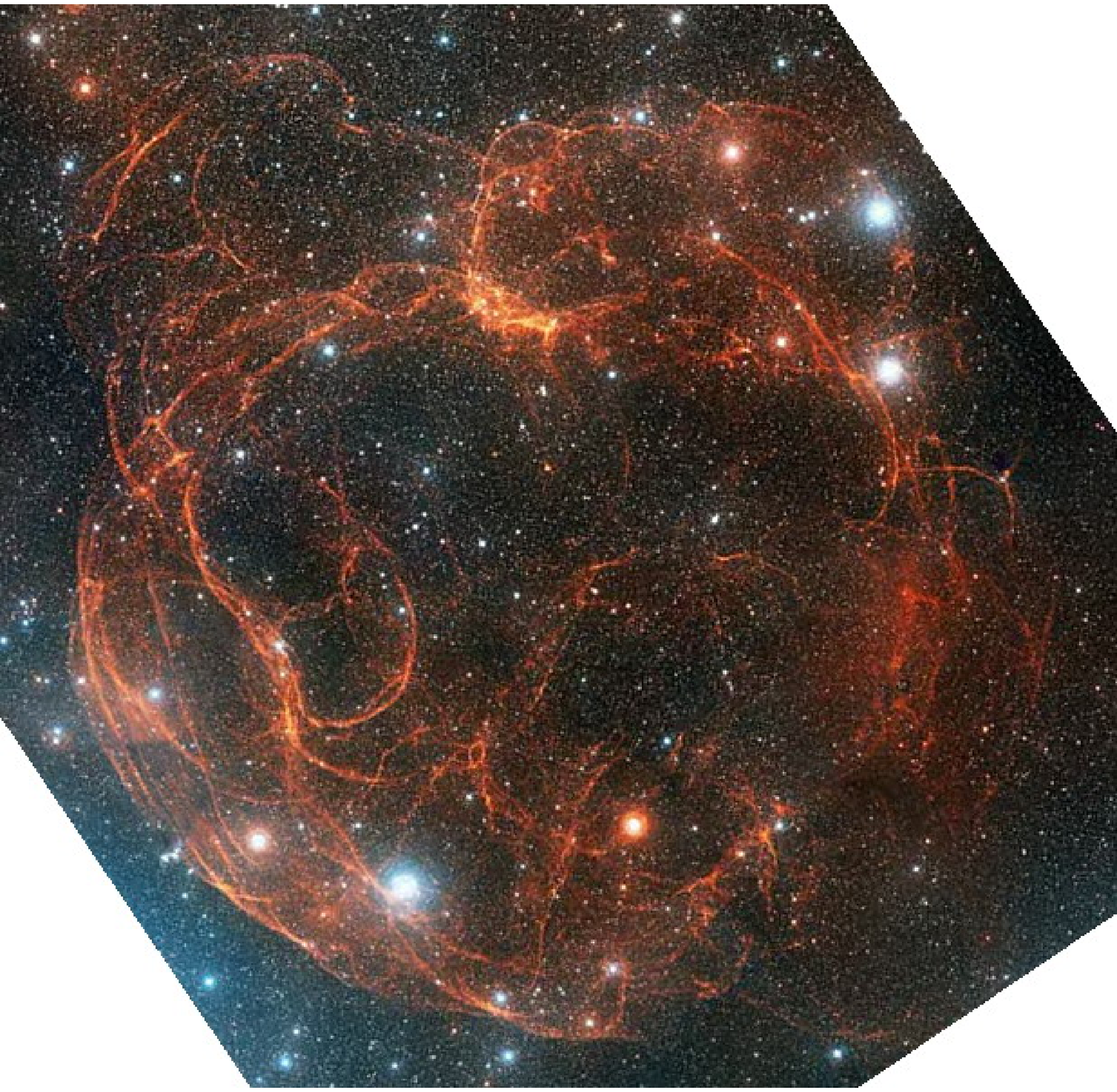}
\includegraphics[width=0.32\textwidth]{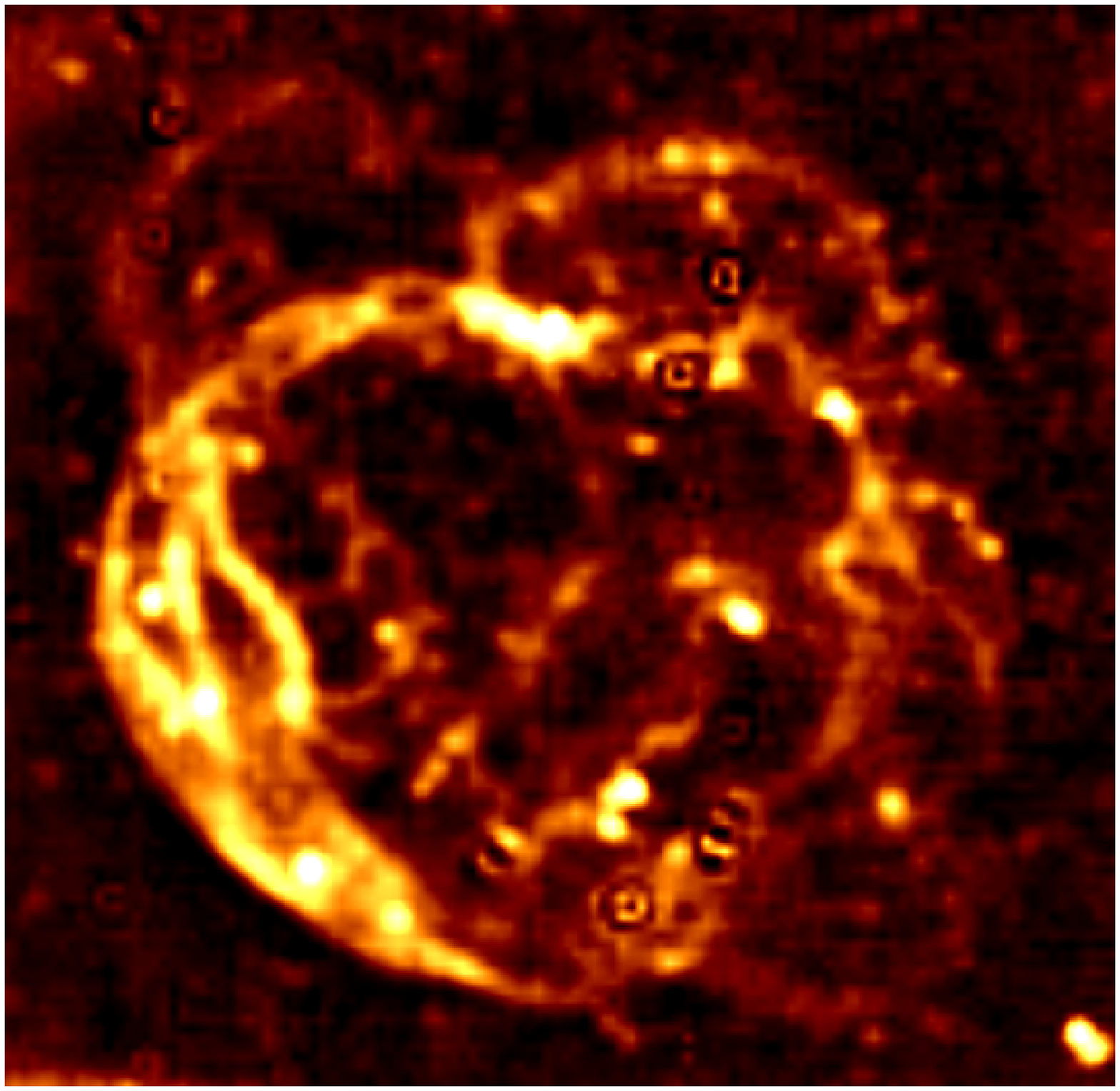}
\includegraphics[width=0.34\textwidth]{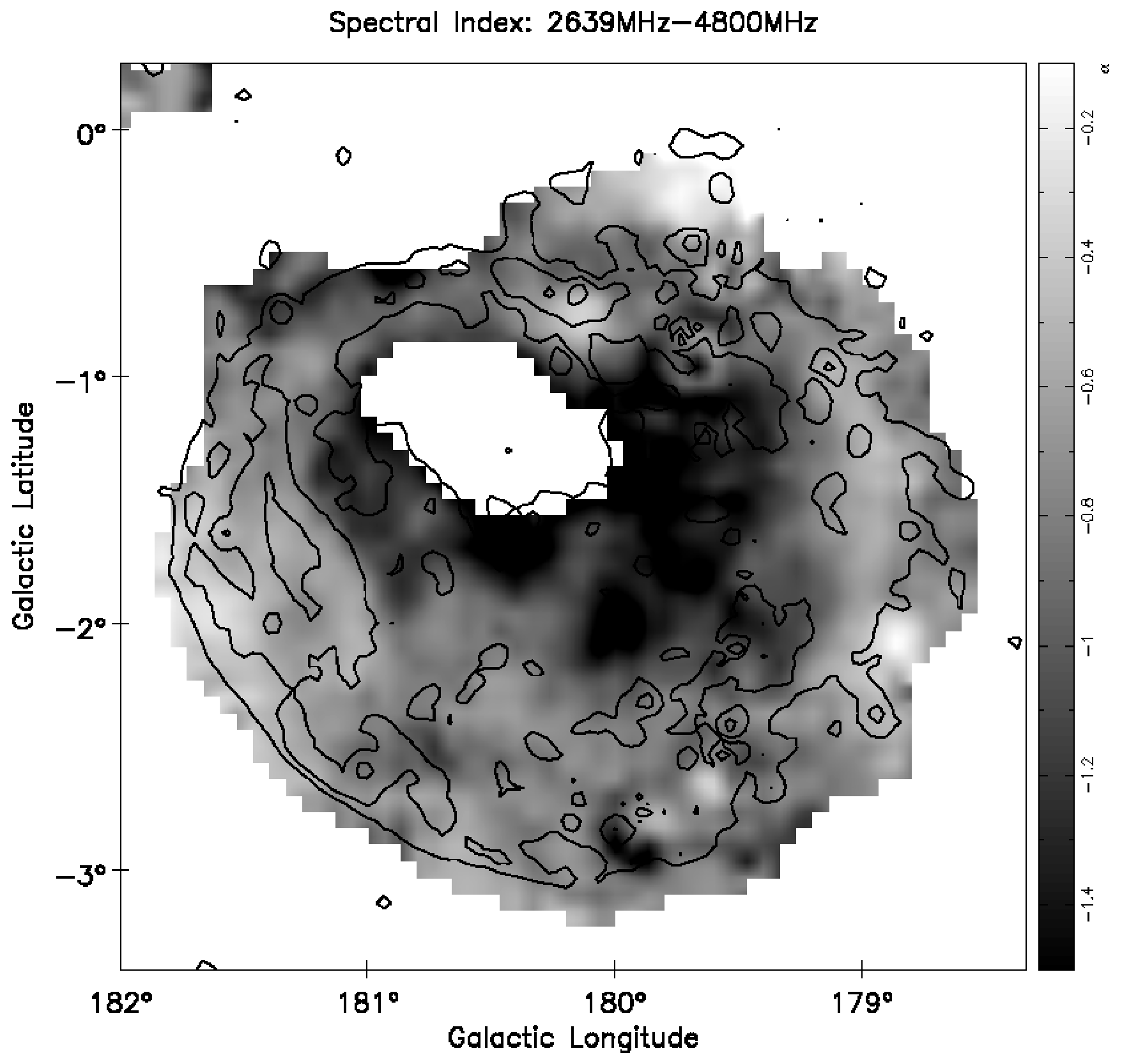}
\caption{The filaments of SNR S147 as observed in the optical
  {(\it left)} and radio ($\lambda$11-cm total power map in the {\it
    middle}). The spectral index map {\it (right)} shows flat
  spectra in the filamentary area, and steep spectra in
  the diffuse area, especially in the central part.} 
\label{s147spec}
\end{figure}
\subsection{Spectral index maps of large SNRs}

Our observations at $\lambda$6~cm are the first to cover many large
SNRs. By combining the maps with these at $\lambda$21~cm and
$\lambda$11~cm observed with the Effelsberg 100~m radio telescope or
other telescopes, we can get their spectral index maps. The first
object is the Cygnus loop shown in Fig.~\ref{cygnus} (\cite{srh+06}).

We obtained spectral index maps for many large SNRs, for example,
G65.2+5.7 (\cite{xrfh09}), G156.2+5.7 (\cite{xhs+07}), S147 in
Fig.~\ref{s147spec} (\cite{xfrh08}), CTA1 (\cite{srw+11c}). These maps
show that the peripheries of SNRs always have a flatter spectrum,
while the diffuse emission in the central part usually has a steeper
spectrum. Physical reason is that newly accelerated particles in the
shock fronts along the SNR periphery radiate in compressed magnetic
fields, and older particles radiate in weaker magnetic fields in other
areas.

\begin{figure}[tb]
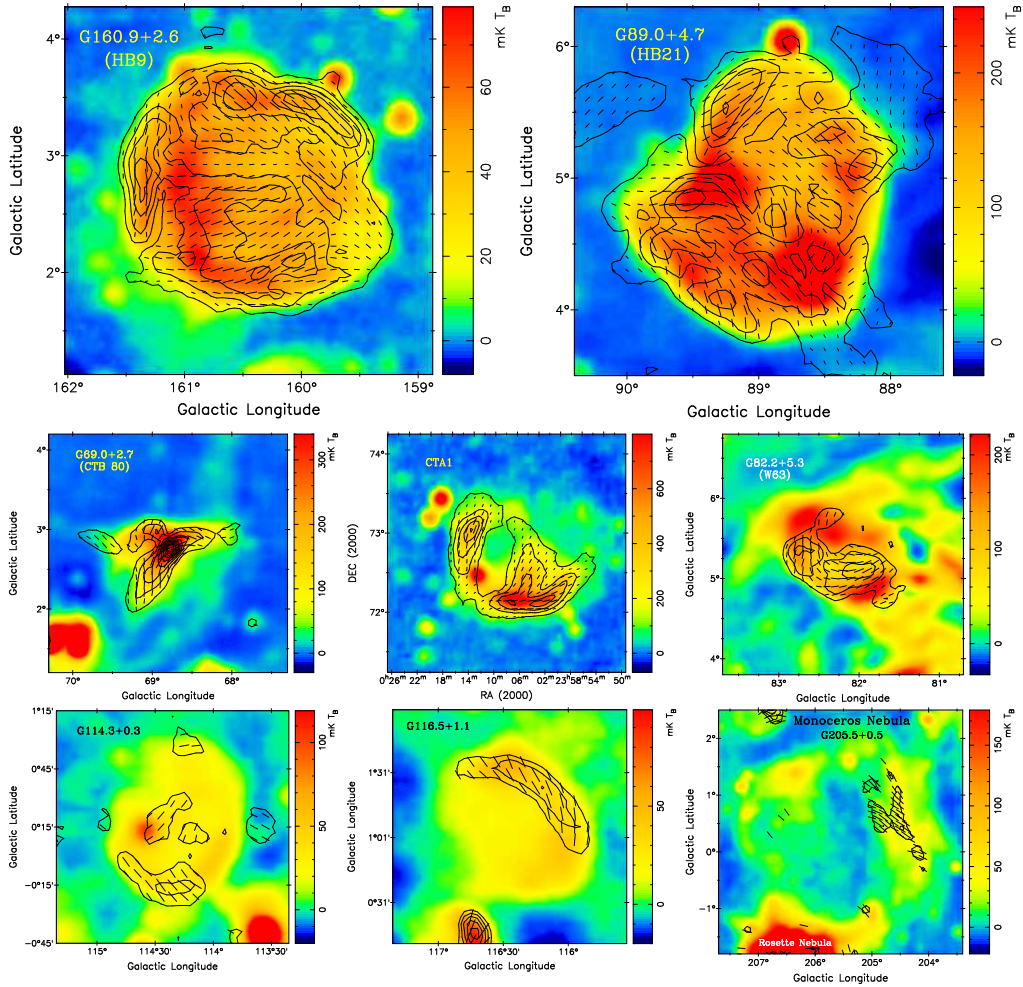

\resizebox{0.495\textwidth}{!}{\includegraphics[angle=-90]{han_fig5a.ps}}
\resizebox{0.495\textwidth}{!}{\includegraphics[angle=-90]{han_fig5b.ps}}\\[1mm]
\resizebox{0.325\textwidth}{!}{\includegraphics[angle=-90]{han_fig5c.ps}}
\resizebox{0.325\textwidth}{!}{\includegraphics[angle=-90]{han_fig5d.ps}}
\resizebox{0.325\textwidth}{!}{\includegraphics[angle=-90]{han_fig5e.ps}}\\
\resizebox{0.325\textwidth}{!}{\includegraphics[angle=-90]{han_fig5f.ps}}
\resizebox{0.325\textwidth}{!}{\includegraphics[angle=-90]{han_fig5g.ps}}
\resizebox{0.325\textwidth}{!}{\includegraphics[angle=-90]{han_fig5h.ps}}
\caption{New polarization maps of some SNRs. Color image for total
  intensity, contours for the polarized intensity and vectors for
  {\bf B} orientation.}
\label{pipa}
\end{figure}
\subsection{New polarization maps}

\begin{figure}[tb]
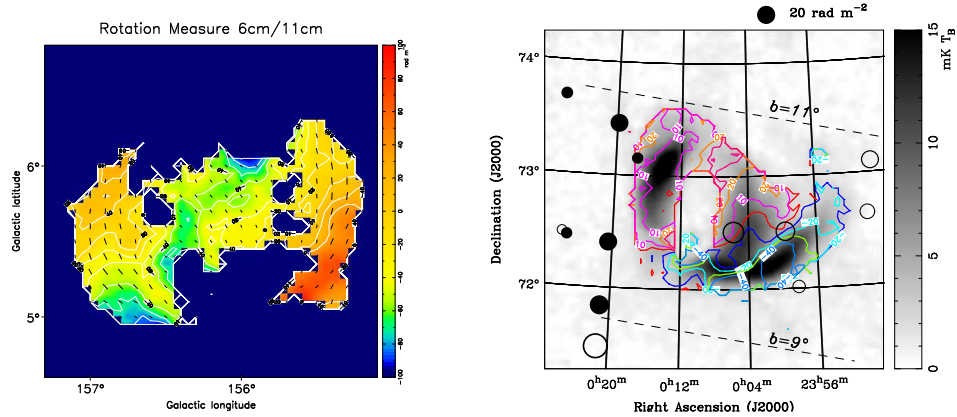
 
\resizebox{0.53\textwidth}{!}{\includegraphics[angle=-90]{han_fig6a.ps}}
\resizebox{0.46\textwidth}{!}{\includegraphics[angle=-90]{han_fig6b.ps}}
\caption{Rotation measure maps of G156.2+5.7 ({\it left}, \cite{xhs+07})
  and CTA1 ({\it right}, \cite{srw+11c}), showing magnetic field structure 
  in SNR or foreground interstellar medium.}
\label{rm}
\end{figure}

At $\lambda$6~cm, due to small angles of foreground Faraday rotation,
we see more or less the intrinsic magnetic field orientations of SNRs from
the observed $\vec{\bf E}+90^{\circ}$. As shown in Fig.~\ref{pipa}, we detect
tangential fields in SNR shells. For some SNRs, we detected very
ordered fields, for example, in the central patch of G156.2+5.7
(Fig.~\ref{rm}, \cite{xhs+07}) and the central branch of CTA~1
(Fig.~\ref{pipa}, \cite{srw+11c}).  Polarized emission is detected
even for the whole SNR area, e.g. in the Cygnus loop in Fig.~\ref{cygnus},
G89.0+4.7 (HB21) and G160.9+2.6 (HB9) in Fig.~\ref{pipa}. We detected
radial magnetic fields in IC443 (\cite{ghr+11a}), and a T-shape
magnetic field in G68.0+2.7 (CTB80 in Fig.~\ref{pipa},
\cite{ghr+11a}) and its very polarized east arm. The polarized
emission of G82.2+5.3 (W63) is found to be anti-correlated with the
radio total power and H$\alpha$ emission (\cite{ghr+11a}), which
indicates a mixture of local thermal and nonthermal emission in the
complex region.

For 23 of 79 observed SNRs, we get the first complete polarization
images at $\lambda$6~cm. A few SNRs were never observed before in
polarization (SNR G205.5+0.5 = Monoceros Nebula; G206.9+2.3;
G85.9$-$0.6, G69.7+1.0, G16.2$-$2.7) until our $\lambda$6~cm
measurements. For G16.2$-$2.7, G69.7+1.0 and G85.9$-$0.6, the
polarized emission is detected for the first time, adding evidence
that they are in fact SNRs (\cite{srr+11b}).

\subsection{Rotation measures in SNRs}

Comparing the polarization angle maps at $\lambda$6~cm from our Urumqi
observations with those of $\lambda$21~cm and $\lambda$11~cm observations 
with the Effelsberg 100~m telescope, we calculated RM maps of SNRs.

Two very interesting examples are shown in Fig.~\ref{rm}. The RM map
of G156.2+5.7 (\cite{xhs+07}) shows decreasing RMs downwards along the
left shell of the SNR and increasing RMs along the right shell. This
indicates a twisted field structure probably in the SNR shell or a
foreground RM gradient. The RM map of CTA~1 (\cite{srw+11c}) show
negative RMs in the southern shell, different from the positive RMs
in the northern shell and central branch, which are caused by a Faraday
screen with reversed magnetic fields in the foreground interstellar
medium. The RM signs are consistent with the RM distribution of
background radio sources in a wider area around CTA~1.

\subsection{Discovery of two large SNRs: G178.2$-$4.2 and G25.1$-$2.3}

After observations for the Sino-German 6cm polarization survey of the
Galactic plane were finished in 2009, extensive data processing was
carefully done by several (former) PhD students. X.~Y. Gao found the
extended source G178.2$-$4.2 and X.~H. Sun found G25.1$-$2.3. Both of
these objects are found to have a nonthermal spectrum. G178.2$-$4.2
has a polarized shell. These two objects have a size of more than
1$^{\circ}$, and are identified as SNRs (\cite{gsh+11b}). See Gao et
al.  in this volume for details.

Our $\lambda$6~cm survey data were also very important to identify 
two new SNRs, G152.4$-$2.1 and G190.9$-$2.2 by \cite{fcr+13}.

\subsection{Disapproved ``SNRs'': OA184, G192.8$-$1.1, G16.8$-$1.1
 and half of the Origem loop}

Using our $\lambda$6~cm data, together with $\lambda$11~cm and
$\lambda$21~cm data from the Effelsberg 100~m telescope, we
disapproved 3.5 ``known SNRs'': OA184 (\cite{fks+06}), G192.8$-$1.1
(\cite{ghr+11a}), G16.8$-$1.1 (\cite{srh+11a}), and the lower half of
the Origem loop (\cite{gh13}).

The first disapproved SNR was OA184 (\cite{fks+06}). T. Foster noticed
its flat spectra at low frequencies, in contrast to the nonthermal
spectra of G166.0+4.3 and HB9 located in the same area of the Galactic
plane. The $\lambda$6~cm observation were added in this investigation
and give the final clue. It turns out that OA184 has a thermal
spectra. More importantly, it appears as a depolarized extended
source in the $\lambda$6~cm map, rather than showing ordered
polarization as expected from a shell-type SNR.

The second disapproved SNR was G192.8$-$1.1 (\cite{ghr+11a}). The
bright knots on the plateau have been found to be either known
background sources or known HII region, plus a newly identified HII
region.  The plateau itself is found to have a thermal spectrum
without any associated polarized emission.

The third one is G16.8$-$1.1 (\cite{srr+11b}). This object in the very
inner Galaxy appears as a depolarization source in our 6cm map
embedded in a large polarization patch of the Galactic diffuse
emission. It coincides well with the known HII region, SH~2-50.

In addition, the Origem loop is found to to be composed of a SNR
arc in the north and HII regions in the South (\cite{gh13}).

\section{Summary}

We have demonstrated that a small telescope is very useful to observe
large objects. By observing large SNRs, of a few degree in size, we
obtained many unique polarization images at $\lambda$6~cm. Many of
these images are the first, or are the one at highest frequency so
far, to reveal the intrinsic magnetic fields of SNRs. Multi-band
observations are very important to calculate spectra of SNRs, or to
get the spectral index images. The observations we present are very
useful for studying the physical properties of SNRs on particle
acceleration and magnetic fields. Using our data, we disapproved three
and half ``known SNRs'' and dismissed the suggested spectral break of
a few SNRs.

Readers can get more information from the web-page: http://zmtt.bao.ac.cn/6cm/.

\begin{acknowledgments}
We thank the SOC of IAUS 296 for their invitation to give this summary
talk, and we acknowledge financial support from the National Natural
Science Foundation of China (10473015, 10773016), specifically for the
Sino-German $\lambda$6~cm polarization survey of the Galactic plane.
The Sino-German cooperation was supported via the partner group of the
MPIfR at the NAOC as part of the exchange program between the MPG and
the CAS for many bilateral visits.
\end{acknowledgments}

\end{document}